\documentclass{an}
\usepackage{times}
\usepackage{graphicx}

\begin{document}
\Pagespan{99}{}% Document's page range. 
% If second parameter is left empty, the last page is computed automatically.
\Yearpublication{2008}%
\Yearsubmission{2008}%
\Month{99}%   
\Volume{999}%  
\Issue{99}% 

\title{How semiregular are irregular variables?}

\author{T. Lebzelter\inst{1}\fnmsep\thanks{Corresponding author:
  \email{lebzelter@astro.univie.ac.at}\newline}
\and M. Obbrugger\inst{1}
}
\titlerunning{Irregular variables}
\authorrunning{T. Lebzelter \& M. Obbrugger}
\institute{
Department of Astronomy, University of Vienna,
T\"urkenschanzstrasse 17, A1180 Vienna, Austria}

\received{}
\accepted{}
\publonline{}
\keywords{stars: AGB and post-AGB -- stars: variables: general -- stars:late type}
\abstract{We investigate the question whether there is a real difference in the light change
between stars classified as semiregular (SRV) or irregular (Lb) variables by analysing
photometric light curves of 12 representatives of each class. Using Fourier analysis we
try to find a periodic signal in each light curve and determine the S/N of this signal.
For all stars, independent of their variability class we detect a period above the 
significance threshold. 
No difference in the measured S/N between the two classes could be found.
We propose that the Lb stars can be seen as an extension of the SRVs towards shorter
periods and smaller amplitudes. This is in agreement with findings from other quantities which also
showed no marked difference between the two classes.} 
\maketitle

\section{Introduction}
Light variability is a common phenomenon among the highly evolved stars on the upper part of the 
giant branch. The variables found there are called long period variables (LPVs), but this class actually
summarizes a number of not very well defined sub classes differing in light amplitude, period and regularity of the light change. 
The classical picture defined in the General Catalogue of Variable Stars (Kholopov et al. 1985--88) 
separates miras, semiregular and irregular variables. For stars with well known distances and 
thus luminosities like objects in the Magellanic Clouds or globular clusters a revised classification
according to the pulsation mode has emerged recently (e.g. Wood et al. 1999, Lebzelter \& Wood 2005).

Among the three classical LPV classes, the irregular variables are clearly the least studied ones.
For the other two classes extensive studies on their properties, also beyond the light change,
are found in the literature (e.g. Kerschbaum \& Hron 1992 and 1994, Jura \& Kleinmann 1992,
Barthes et al. 1999). The few works that focused on irregular variables (Lb),
however, strongly indicate that these variables cannot really be separated from the other LPVs
in terms of near infared red colours (Kerschbaum et al. 1996), spectral energy distribution 
(Kerschbaum 1999) and mass loss properties (Olofsson et al. 2002). The class of irregular variables
may contain "a slightly larger contamination with non AGB-giants" than the semiregular variables 
(Kerschbaum et al. 1996).

What remains is a difference in the regularity of the light change. Irregular variables are typically
stars for which no sign of periodicity could be found. This, however, could be either due to an intrinsic
irregularity in the variation or due to a bad sampling of the light curve used for the classification.
Based on the latter argument Lebzelter et al. (1995) suspected that a lot of semiregular variables
were erroneously classified as irregular variables. This is supported by the fact that for a number of
stars originally classified as irregular variables a periodicity could later be found from better light
curve data or an improved data reduction (e.g. Percy et al. 1996). 

In this paper we analyse and compare the light curve characteristics of samples of semiregular and irregular 
variables in order to test if there is a real difference in the light curve characteristics of these two
variable star classes. The length of our datasets, typically around 1000 days, extends significantly beyond what has been used to classify most of the long period variables in the General Catalogue
of Variable Stars (compare Lebzelter et al. 1995). 

\section{Sample}
For our study we used a sample of 12 light curves of irregular variables (type Lb) obtained with
the Austrian automatic photoelectric telescope (APT) located in Southern Arizona. The classification as 
irregular variables was taken from the General Catalogue of Variable Stars
(Kholopov et al. 1985--88). Targets were selected only on their brightness (optimized for the
instrument) and their location in the sky to allow for a long visibility from Arizona.
The telescope and the
photometric instrument are described in detail in Strassmeier et al. (1997). Data were obtained
in Johnson $V$ and $I_{C}$. The time series were part of a larger
study on long period variables (e.g. Lebzelter 1999, Kerschbaum et al. 2001). Table \ref{Lbsample}
gives the basic parameters of the obtained time series.

\begin{table*}
\begin{center}
\caption{Basic light curve parameters of our Lb sample. The second column gives the
total number of data points of the investigated light curve. Column 3 lists the
time span, during which these observations were taken (JD$-$2\,400\,000). 
Column 4 gives the detected primary period of the light change. 
The value in brackets is the S/N measured for the corresponding period (see text for details).
Column 5 gives the amplitude resulting from the Fourier analysis, and column 6
lists the maximum observed light amplitude (min $-$ max). All amplitudes are
given in magnitudes peak-to-peak.}  \label{Lbsample}
\begin{tabular}{lccccc} 
\hline
Name & Number of  & Time Span of & Periods (S/N) & Amplitude & Total Range\\
     & Datapoints & Light Curve  &               &           & of Light Change\\
\hline
AA Cas & 211 & 50711--51519 & 82 (14.3) & 0.30 & 0.64\\
V1125 Cyg & 306 & 50564--52177 & 82 (12.1) & 0.22 & 0.58\\
V1351 Cyg & 233 & 50564--51517 & 113 (13.9) & 0.22 & 0.46\\
FK Hya & 426 & 50492--52030 & 101 (12) & 0.24 & 0.91\\
FZ Hya & 509 & 50477--52642 & 136 (12.2) & 0.36 & 1.08\\
FZ Lib & 443 & 50473--52087 & 26 (6.5) & 0.06 & 0.39\\
V398 Lyr & 329 & 50564--52179 & 109 (9.4) & 0.22 & 0.82\\
GO Peg & 173 & 50721--51528 & 72 (9) & 0.21 & 0.54\\
DV Tau & 83 & 50466--51212 & 62 (4.7) & 0.22 & 0.97\\
AZ UMa & 521 & 50464--52076 & 212 (20) & 0.46 & 1.25\\
TT UMa & 473 & 50465--52051 & 62 (8.9) & 0.20 & 1.09\\
RW Vir & 451 & 50493--52075 & 116 (8.5) & 0.18 & 1.76\\
\hline
\end{tabular}
\end{center}
\end{table*}

We briefly summarize the relevant publications on the various targets from our sample of irregular
variables restricted to studies of the light variability. The light curve of one star, {\it RW Vir}, has been presented and discussed already in 
Kerschbaum et al. (2001). Five of the targets were included in a radial velocity study by Lebzelter
\& Hinkle (2002), namely {\it V1351 Cyg, V398 Lyr, FK Hya, FZ Hya} and {\it RW Vir}. In that paper
the same dataset was used to estimate the pulsation phase of the stars. Lebzelter \& Hinkle did not undertake
a detailed light curve analysis, but gave typical values for the timescale of variability derived
from a Fourier analysis. We made an independent re-analysis of all the data for this paper and found
the same main period (within one or two days) in all cases except for {\it V1351 Cyg}, where
Lebzelter \& Hinkle give a period of 44 days. 

In general, very little has been published on the light variability of these stars. For two of the remaining 
11 stars attempts to derive a period were published ( i.e. {\it AA Cas} and {\it V1351 Cyg}). 

Houk (1963) included {\it AA Cas} in the list of
long period variables with a secondary period giving a range of 70 to 115 days as the first period and
850 days as the second. 
Foster (1995) made a fit of the AAVSO light curve of this star using 7 periods (1790, 700, 380, 121, 80.25,
76.1 and 71.85 days, respectively). Four years later the star was included in the detailed period search in AAVSO data carried out by Kiss et al. (1999), but no periodicity could be detected. Adelmann (2001) measured an amplitude from Hipparcos data of 0.79\,mag.\\
A light curve of approximately 70\,d length of {\it V1351 Cyg} was analysed by Percy \& Fleming
(1992). They note that the timescale of the variability is 120\,d or more with an amplitude of
0.45\,mag or higher in $V$. The Hipparcos light curve of this star was inspected by Koen \& Eyer (2002). They detected a period of 69\,d with an amplitude of about 0.1\,mag.\\
The most quaint story relates to {\it FZ Lib}. This star
was used as a standard star by Stokes (1971). Later Eggen (1976) noted its variability, but his few
measurements did not allow to describe the light change. 

To see if the light change of irregular variables is significantly different from the light
change of semiregular variables we compare them with a sample of light curves of 
semiregular variables obtained with the same
instrument. A period analysis of most of these objects has been published already elsewhere
(Kerschbaum et al. 2001, Lebzelter \& Kiss 2001, Lebzelter \& Hinkle 2002), 
but to avoid systematic errors we did a re-analysis
of these stars with the same method as the irregular variables. The randomly 
selected 12 semirgeular variables are listed in
Table \ref{SRVsample}. The period found by Lebzelter \& Hinkle (2002) is only given in cases
where their analysis lead to a period strongly different from the GCVS value (see their paper for
details). 

\begin{table*}
\begin{center} 
\caption{Basic light curve parameters of our SRV sample. See Table \ref{Lbsample}.
Column 5 gives the GCVS period for comparison (taken from the online version of
the GCVS at http://www.sai.msu.su/groups/cluster/gcvs/cgi-bin/search.htm). 
Column 6 gives the first periods found on the same dataset before (LK...Lebzelter \& Kiss 2001,
KLL...Kerschbaum et al. 2001; LH...Lebzelter \& Hinkle 2002; 
long time trends excluded). All periods are given in days. The final two columns
give the light amplitude (see Tab.1).} \label{SRVsample}
\begin{tabular}{lccccccc} 
\hline
Name & Number of  & Time Span of & Periods (S/N) & GCVS & earlier & Amplitude & Total Range\\
     & Datapoints & Light Curve  & Period        &      & papers  &           & of Light Change\\
\hline
EP Aqr & 202 & 50713--52181 & 184 (11.1) & 55 & & 0.34 & 0.82\\
RV Cam & 334 & 50492--51981 & 176 (16.5) & 101 & 177, 109 (KLL) & 0.54 & 1.29\\
RY CrB & 386 & 50466--52438 & 87 (10.4) & 90 & & 0.40 & 1.26\\
U Del & 453 & 50518--52603 & 117 (6.7) & 110 & 174, 205 (LK), 205 (LH) & 0.24 & 1.05\\
g Her & 480 & 50520--52643 & 89 (10.8) & 89 & 89 (KLL), 89 (LK) & 0.20 & 0.81\\
UW Her & 421 & 50467--52177 & 106 (20.2) & 104 & & 0.68 & 1.50\\
X Her & 409 & 50504--52087 & 100 (15) & 95 & 101, 165 (LK) & 0.34 & 0.84\\
EY Hya & 157 & 51529--52029 & 173 (16.6) & 183 & & 0.54 & 0.90\\
SX Leo & 412 & 50468--52060 & 122 (13.3) & 100 & & 0.26 & 0.78\\
FY Lib & 352 & 50482--52087 & 102 (11) & 120 & 103 (LH) & 0.26 & 0.80\\
SY Lyr & 403 & 50494--52085 & 163 (14.4) & 100 & & 0.38 & 1.04\\
BK Vir & 182 & 51530--52076 & 146 (22.1) & 150 & & 0.70 & 0.93\\
\hline
\end{tabular}
\end{center}
\end{table*}

\begin{figure}
\includegraphics[width=80mm,height=69mm,clip]{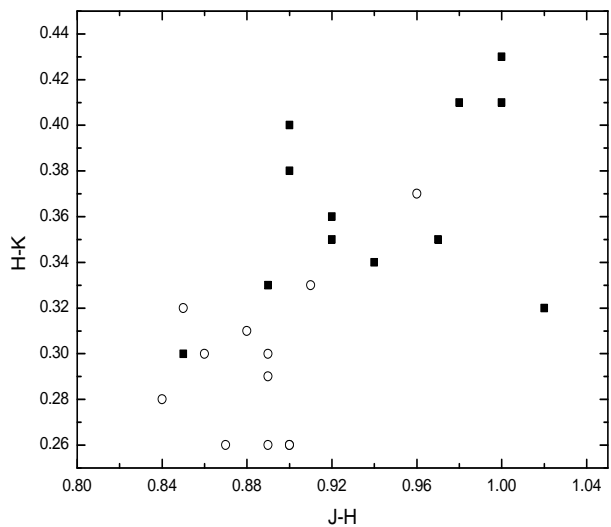}
\caption{$J-H$ vs. $H-K$ diagram for our sample of semiregular (filled boxes) and
irregular (open circles) stars. A typical error bar (according to Kerschbaum \& Hron 1994)
is given at the lower right corner.}
\label{nircol}
\end{figure}

To better characterize our samples we extracted $J-H$ and $H-K$ colours from the literature
(Kerschbaum et al. 1996, Kerschbaum \& Hron 1994). In Figure \,\ref{nircol} we give a
two colour plot for both samples. The uncertainty of the colour values is typically 0$\fm$04
(see Kerschbaum \& Hron 1994).
According to the classification of the SRVs
by Kerschbaum \& Hron (1994) we find that most of our semiregular stars belong to the group
of 'red' SRVs. The irregular variables are predominantly found in the area of the
'blue' SRVs in agreement with the findings from Kerschbaum \& Hron. Our two samples
show a clear overlapping region in this two colour diagram.

\section{Light curve analysis}

We start with some considerations on the light curve ana\-lysis. First, we
will restrict our discussion to a Fourier ana\-lysis approach as it is still the most commonly used one
for long period variables
(e.g. Kiss et al. 1999), although other approaches can be also found in the literature
(e.g. Eyer \& Genton 1999). In the Fourier analysis, or more specifically in the
Discrete Fourier Transform, of astronomical data sets the primary interest is to
find a couple of eigenfrequencies in order to describe the general variability pattern
rather than a complete signal recovery. As described in Reegen (2007) the typcial strategy
is to identify the local maxima in the amplitude spectrum and to determine their significance.
Further significant periods may be detected by subtracting the detected sinusoidal signal
from the data set. A fine tuning of detected frequencies can be done using a least square
fitting technique (e.g. Sperl 1998).

We are therefore using the amplitude spectra of our sample light curves for the comparison of
the two variable classes.
The frequency analysis was performed using SigSpec (Reegen, 2007) and PERIOD04 (Lenz \& Breger, 2005). Since both methods produce very similar results, we present the spectral analysis conducted using PERIOD04 only. Both programs give a signal-to-noise (S/N) value for each detected peak in the amplitude spectrum. 
The S/N ratios of the resulting frequencies were calculated after prewhitening the corresponding frequency. 
The applied criterion for the significance of the frequencies was a S/N ratio higher than 4.0 (Breger et al. 1993 and Kuschnig et al. 1997). We refer to the respective papers on the two programs used for a description of how the S/N values were determined.
Periods obviously caused by observational gaps in the data set were excluded by visual inspection
of the lightcurves and the fits. In order to extract the most significant intrinsic frequency the data sets were corrected for such peaks before doing the further analysis.
As presented in Table \ref{Lbsample} and Table \ref{SRVsample} 
all signals were well resolved (i.e. within the boundaries given by the length of the data set and their respective Nyquist frequency (Nyquist 1928)).

\begin{figure}
\includegraphics[width=80mm,height=45mm,clip]{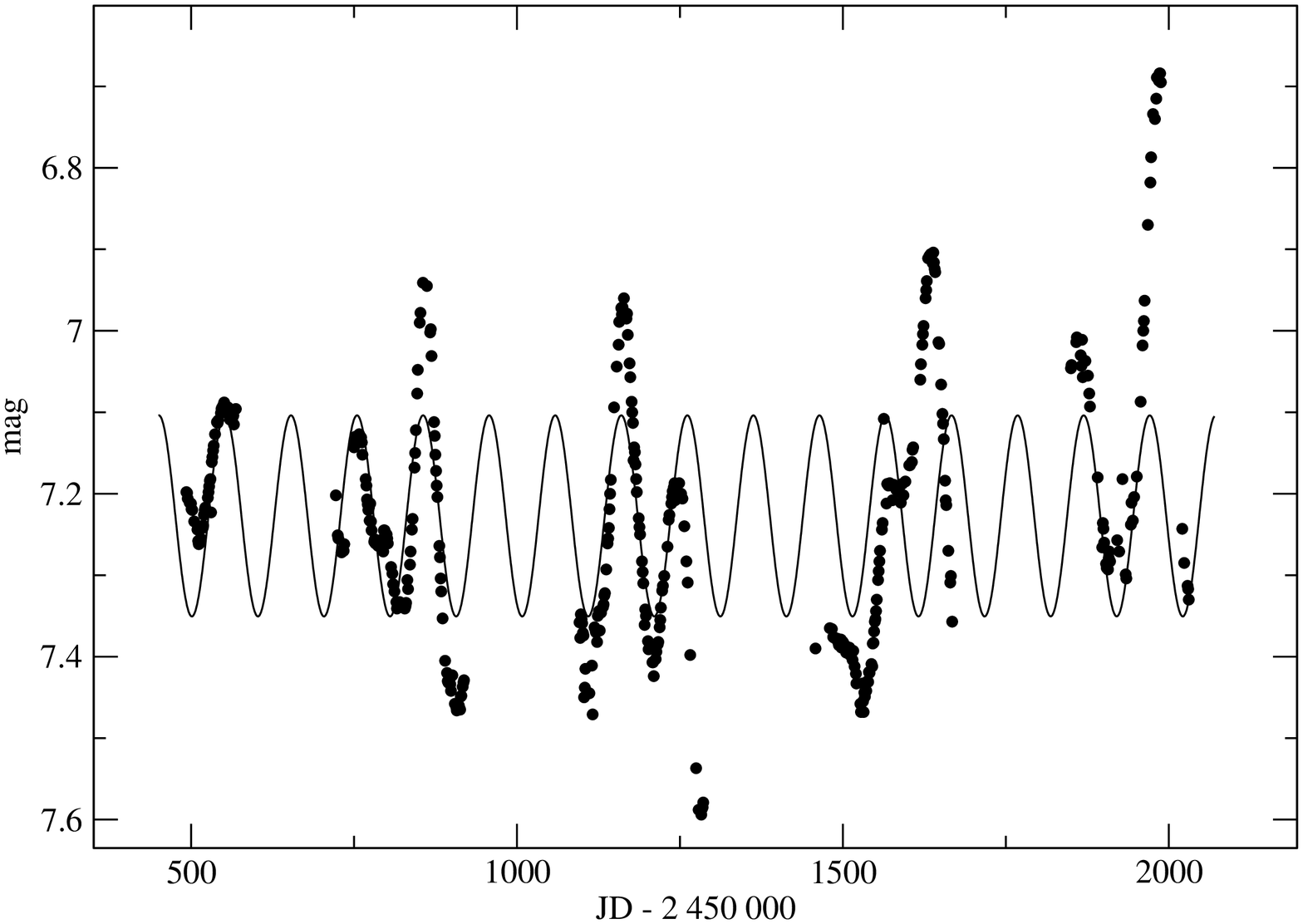}
\hfill
\includegraphics[width=80mm,height=45mm,clip]{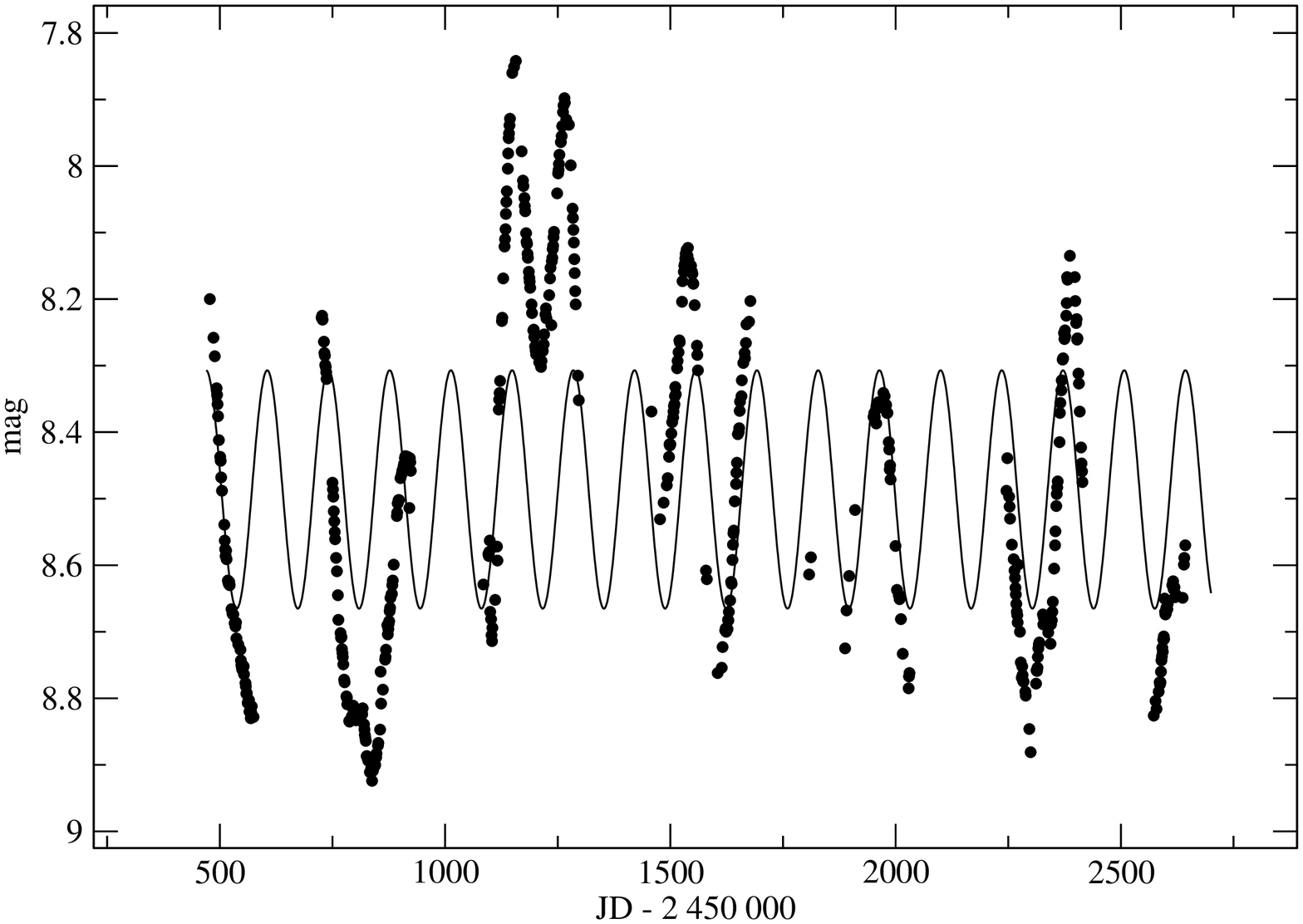}
\caption{$V$-band light curves of FK Hya (upper panel) and FZ Hya (lower panel). Also
shown are Fourier fits with the primary period detected. The photometric
accuracy of the individual data points is typically better than 0$\fm$02.}
\label{lc_example}
\end{figure}

In Figure \ref{lc_example} we show the light curves of two irregular variables of our sample together with
a fit with the primary period detected. The resulting periods found for the irregular variables
are listed in Table \,\ref{Lbsample} together with their S/N ratio. The same numbers are given
for the semiregular variables in Table \,\ref{SRVsample}. In both tables the periods derived
from the $V$ band and the amplitude resulting from the corresponding Fourier fit are listed. We also give the total $V$-band light amplitude
in the last column.

To estimate the frequency uncertainty for our Fourier analysis we can first consider the resolution
achievable in frequency which is indirect proportional to the sample length
(Loumos \& Deeming 1978). For our data set this corresponds to a period uncertainty of 10 to 30\%
(see Kerschbaum et al. 2001). However, this conservative error is expected to be significantly reduced by the also applied least square fitting in PERIOD04. Kallinger et al. (2008) showed that 
for a monoperiodic signal the uncertainty in frequency is in this case only about $0.3/T$ for
a S/N ratio around 10. This would correspond to period uncertainties of 5\% or less even 
in multiperiodic case (cf. Kallinger et al. 2008). In the present case we have to consider that
the signal detected is probably not strictly periodic. 
 
As mentioned above the main period found for {\it RW Vir} is in agreement with the earlier analysis of
this dataset (Kerschbaum et al. 2001). Taking into account the uncertainties of the literature values
for the period of {\it AA Cas} we see a good agreement with our findings for this star. 
On the opposite, the 69\,d period of {\it V1351 Cyg} reported by Koen \& Eyer (2002) could not be 
confirmed by our data. For this star we have one period of 113 days from our data which is in 
agreement with the estimate from Percy \& Fleming (1992). We see a second period of 80 days
which has to be compared with 44 d (Lebzelter \& Hinkle 2002) and the just mentioned 69 d period.
We conclude that there is probably a variation on a timescale of a few ten days, but it clearly needs further
investigation.

When comparing our results for the SRVs with previously published values, we have to keep in mind
that several of the earlier studies aimed to reproduce the light change very well with a set of
up to six frequencies. Optimizing the fit was done in parallel for all these sequences (see e.g. 
Lebzelter \& Kiss 2001). Thus it has to be expected that the resulting primary frequency differs from
the value found by us, as we were not aiming to make a perfect fit and instead are only interested
in the strongest peak showing up in the Fourier analysis. 
Lebzelter \& Kiss (2001) noted
further by comparing their results with findings from other authors that the multiperiodic fit is
not unique for several of the stars investigated by them. 

In the light of these considerations we can now compare our findings for the semiregular stars with
the published periods. Seven stars of our sample show good agreement with the GCVS value.
For {\it RV Cam} our results are supported by the findings from Kerschbaum et al. (2001). These authors
found the GCVS period as a second frequency in their analysis. Our analysis of {\it FY Lib}
gives a shorter period than the one listed in the GCVS, but our value agrees with the findings from
Lebzelter \& Hinkle (2002). Note that in both cases only the analysis can be compared as
the same data sets were used. {\it U Del}, on the other hand, shows in our analysis a period close to
the GCVS value, but in disagreement with two earlier analyses of the same data. To our understanding
the reason for this difference lies in the handling of the long period variation in this star, which
can be modelled in various ways. For {\it SY Lyr} the GCVS period can be found in the Fourier analysis,
but with a smaller amplitude than the 163 d period listed here. The case is different for {\it EP Aqr}
as the 55 d GCVS period is not seen in our data. In general, the light change is not well reproduced by one or two frequencies for this star. Lebzelter \& Hinkle (2002) thus chose the GCVS value in their 
analysis. We stress that longer time series may reveal other periods than the ones listed here as
the primary ones. 

The $V$ and $I$ band light curves of the stars were analysed separately and lead to the same
periods within $\pm$2\,d (corresponding to a difference of 1-2\%). 
The only exception is {\it EP Aqr}, where the analysis of the $V$ light curve lead to 184\,d, while the 
$I$ band light curve gave 178\,d. In {\it DV Tau} the period found from the $V$ band was detected
also in the $I$ band, but with a S/N ratio below the significance level. 

As described already in Kerschbaum et al. (2001) the light change in $I$
does not always exactly follow the light change in $V$. However, as seen from the comparison of the
derived periods, the underlying periodicity seems to be the same. In general we find a well expressed
relation between $V$ and $V-I$. This is expected, if the light change results from stellar radial
pulsation (e.g. Gautschy 1987). 

We also compared the $V$ and $I$ data sets concerning the amplitudes of the periods coming out of
the Fourier analysis. For all but one star ({\it V398 Lyr}) the $V$ band amplitude is larger than the
$I$ band amplitude, typically by a factor of 1.5. There is no obvious difference between Lbs and SRVs.
%However, there is a difference between the $V$ and $I$ band data 
%in the S/N ratios of the detected periods, the irregular variables on the average showing a 20\% higher
%S/N ratio of the periods found from the $V$ band than from the $I$ band data.

\section{Discussion}
Two quantities derived from the amplitude spectrum will be used for comparing semiregular and
irregular variables: the primary parameter will be the S/N ratio of the strongest peak in
the amplitude spectrum. In the case of a truely irregular light change we would expect to
see only noise, thus the criterion for a peak to be significant will not be fulfilled. 
The higher the S/N ratio of a peak the lower will be the False Alarm Probability, indicating
more clearly the presence of a real periodic signal (or -- in a broader sense --  a typical
timescale of the variations). 

\begin{figure}
\includegraphics[width=80mm,height=69mm,clip]{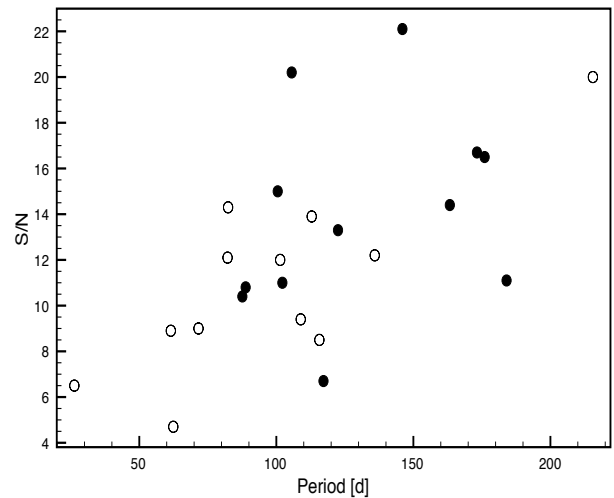}
\caption{S/N ratio of the detected main period against period. Long term changes were excluded. Open
symbols mark irregular variables while filled symbols indicate SRVs.}
\label{SN_comp}
\end{figure}

In Figure \,\ref{SN_comp} we show the S/N ratios of the detected periods ($V$ band) against the period
 for the two classes of variables. Only the period with the highest S/N ratio was plotted excluding long time variations. A general tendency for a higher S/N ratio with longer period can be
seen. The longer periods typically also show a larger total light amplitude (excluding again long time
variations). The semiregular and irregular variables occupy slightly different areas in the diagram, the SRVs
being located at longer periods. Where the two variable groups overlap in period
no difference in the  S/N ratio can be seen. A period (or typical timescale) can be attributed to both kinds of stars with the same
significance. 

However, as pointed out by van Leeuwen (2007a), the uncertainty of the period should be
proportional to the square of the period, which is equivalent to the S/N
for the period to be proportional to the inverse of the period. This means that,
considering the similar time spans observed for all stars, we would expect a trend opposite
to the one seen in Fig.\,\ref{SN_comp}. As revealed by a more detailed analysis the observed
behaviour results from an increase of the S/N ratio with the ratio between the amplitude of 
the first Fourier fit and the total observed light amplitude. This ratio shows
an increasing trend with period, too, which leads to the observed relation between S/N and period.
There is also a general, but not strict trend of an increasing total light amplitude with
period.

The ratio between the amplitude of the first Fourier fit and the total observed light amplitude
gives an indication for the fraction of the light change that can be modelled by a single
sinusoidal fit. For our samples of irregular and semiregular variables the average ratio is 31\,\% 
and 41\%, respectively.
It seems that the possibility to fit the light curve with a single period is increases towards
longer periods. However, this may also be affected by other parameters like the asymmetry of
the light change.

\begin{figure}
\includegraphics[width=80mm,height=69mm,clip]{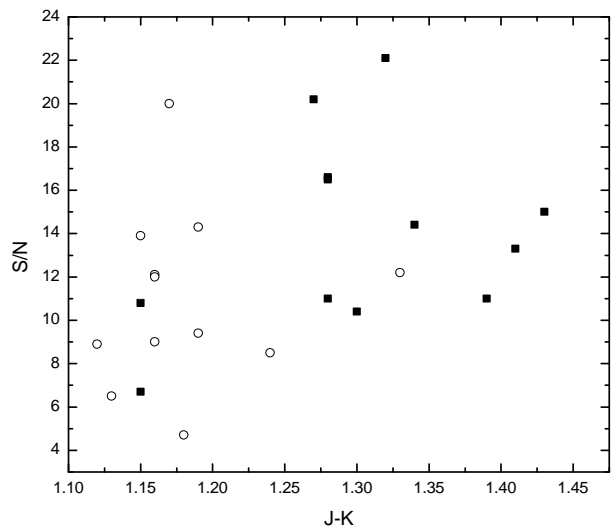}
\caption{S/N ratio of the detected main period against $J-K$. Open
symbols mark irregular variables while filled symbols indicate SRVs.
The irregular variable which is somehow offset from the others is {\it AZ UMa}.}
\label{SN_JK}
\end{figure}

In Figure \,\ref{SN_JK} we compare the near infrared colour $J-K$, which can be seen as an indicator for
effective temperature of the star with the S/N ratio. Combining both samples we see a trend towards
higher S/N with higher values of $J-K$, i.e. lower temperature. As in Figure \,\ref{SN_comp} the irregular
variables form the lower part of the sequence. At the same colour both groups lead
to similar S/N ratios for the primary period. 

For a further comparison of the variability pattern shown by stars from both classes of
variables we investigate if there is a difference between the overall shape of the
amplitude spectrum. In their analysis of a large set of light curves from the AGAPEROS
survey Lebzelter et al. (2004) still followed the original classification and divided the
long period variables into regular, semiregular and irregular variables by eye. They suspected
that the Fourier amplitude spectrum should reflect a decreasing regularity by an increasing
number of peaks of similar strength. Lebzelter et al. (2004) tried to quantify this effect
by forming the ratios between the amplitude of the strongest and the second strongest peak 
(a1/a2) and the
second and the fifth strongest peak (a2/a5), respectively (see their Figure  3). 
The first number takes into the account the widely found fact of multiple periods
in semiregular variables. For single periodic
stars this value should be close to the ratio between the main peak and the first alias seen in
the spectral window.

If the second ratio is small, it indicates that the Fourier amplitude spectrum shows a large
number of peaks of similar height. For the typical S/N values we found this means that 
most of these peaks are above the significance level (or aliases). Lebzelter et al. (2004) showed
that these two ratios indeed allow to describe the amplitude spectrum. However, all the irregular and
part of the semiregular variables occupy the same a1/a2 and a2/a5 parameter range, 
i.e. no difference between the two groups was found by their study.

We computed the two ratios for all the stars in our sample. For this step we first subtracted
long periods and obviously erreneous peaks due to the gaps in our light curves. We find that
our SRVs and Lbs are found in the same part of the a1/a2 vs. a2/a5 diagram as the semiregular and
irregular variables of Lebzelter et al. (2004). Most of our stars are found close to the bottom, i.e. at rather
low a2/a5 values. For all but four stars ({\it RV Cam, GO Peg, AA Cas} and {\it FZ Lib}) the ratio a1/a2
is much smaller than the ratio between the highest peak and the first alias peak in the 
respective spectral window diagram. The remaining four stars all have a low a2/a5 value.
Combining these two findings we conclude that in all stars at least one additional
frequency is found with an amplitude similar to the primary frequency. Thus for both
groups of variables the light change cannot be completely described by a single period.

\begin{figure}
\includegraphics[width=80mm,height=69mm,clip]{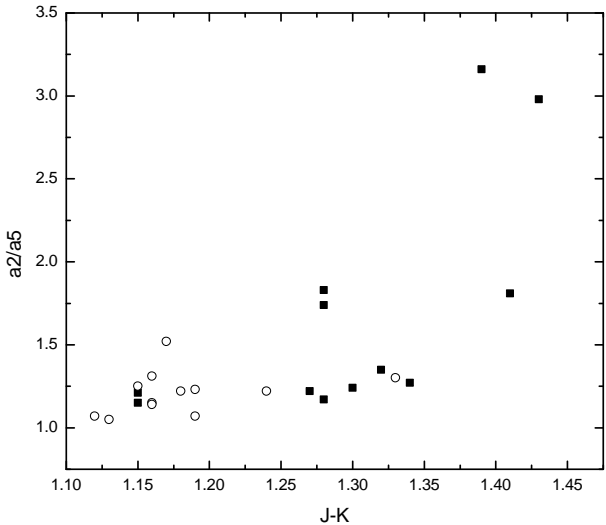}
\caption{Ratio of the second highest and the fifth highest peak in the Fourier amplitude
diagram (a2/a5) vs. $J-K$. See text for details. Same symbols as in Figure \,\ref{SN_JK}.}
\label{a2a5_JK}
\end{figure}

Figure \,\ref{a2a5_JK} shows the ratio a2/a5 against $J-K$. Our data indicate that there is
a steep increase of this ratio towards cooler stars. This would signalize that the Fourier
amplitude diagram becomes more 'clear' for these cooler stars, i.e. the stars tend towards 
a single or double periodic behaviour with lower temperature. This conclusion is supported by the
number of significant peaks in the Fourier amplitude diagram which is decreasing
from an average value of 13 at $J-K=$1.15 to an average value of 4 at $J-K=$1.4.
As in the previous comparisons 
(hotter) semiregular and irregular variables occupy the same region in the diagram.

It is interesting to compare the period distribution in Figure \,\ref{SN_comp} with the overall period distribution
of the stars classified as SRa or SRb in the GCVS. Such a period distribution has been presented
in the outstanding study by Kerschbaum \& Hron (1992). According to their results
our sample of semiregular variables roughly covers the main period range of SRVs and can thus be
seen as representative. Furthermore, the period distribution of SRVs in the GCVS has a steep drop
for periods below 80 days (compare Figure  3 of Kerschbaum \& Hron 1992). This is exactly the period
range where we find a large fraction of our 'irregular' stars. Therefore we can draw the simple 
conclusion that the Lbs actually represent the short period extension of the SRbs. This is in agreement
with their similarity in other stellar parameters (Kerschbaum et al. 1996). 

\begin{figure}
\includegraphics[width=80mm,height=80mm]{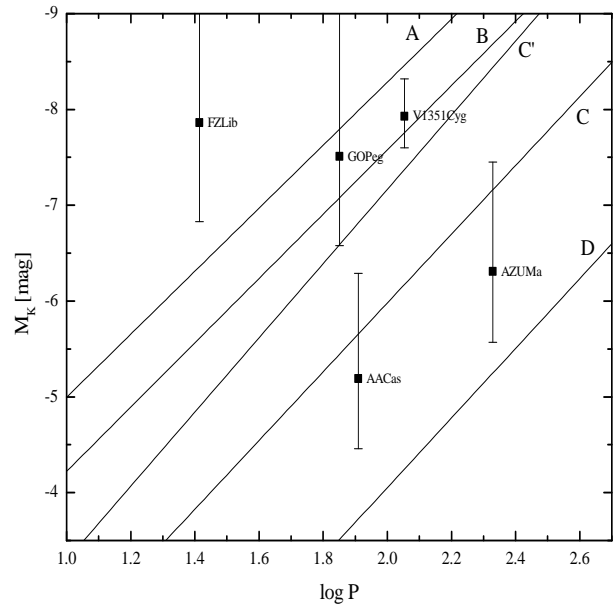}
\caption{Period-luminosity diagram showing the location of 5 Lb variables of our sample with
Hipparcos distances relative to the LMC P-L-relations found by Ita et al. (2004). Each star
is plotted with its primary period only.}
\label{PL}
\end{figure}

For five Lb variables of our sample the Hipparcos catalogue (van Leeuwen 2007b) gives a parallax
where the error of the parallax is smaller than its value. We used these parallax
values to calculate $M_{K}$ values for these stars in order to place them onto a period-luminosity
diagram. In Figure \ref{PL} the result is compared with the location of the various P-L-relations 
found in the LMC (Ita et al. 2004)\footnote{The luminosity relations were shifted from the LMC
distance to absolute $K$ magnitudes by using a distance modulus of 18.5.}. The $K$ values were
taken from 2MASS (Skrutskie et al. 2006). 

Although the error bars are quite large we can at least attribute one star, {\it V1351 Cyg}, to
a pulsation sequence, namely B (second overtone), with its primary period. 
The other stars cannot be attributed to one
sequence that clearly, but obviously the detected periods in {\it FZ Lib} and {\it GO Peg} belong to
overtone pulsation, while the primary period of {\it AA Cas} may be fundamental mode (sequence C).
For the 5 stars in Figure \ref{PL} we searched for a second significant period in the available
light curve data after subtraction of the first signal detected. The second period 
found for {\it V1351 Cyg} (80 days) almost perfectly fits onto sequence A (higher overtone).
However, the second period of {\it AA Cas} (129 days) falls inbetween sequence C and D. 
{\it AZ UMa} also cannot be attributed to a pulsation sequence clearly. A second period of
119 days (S/N=9.89 for the $V$-band data) of that star brings it close to the 
fundamental mode sequence. Finally, also twice the primary period was found in the data.
The analysis of the light curve of {\it GO Peg} gives 106 days as a second period with almost
the same S/N as the primary period. Taking into account the error in the star's distance
it is difficult to say to which sequence this period may belong. For {\it FZ Lib} the 
second highest peak found corresponds to twice the primary period.

We have to note that due to the lack of near infrared light curves we cannot
exclude an additional uncertainty due to the variability of the stars. 
Looking at the observed amplitudes
in the visual range, however, we may safely assume that this effect will be small for most of the
stars studied here. 
Whitelock et al. (2000) report $K$ amplitudes for various long period variables including a 
few semiregular and irregular variables. The amplitudes observed for stars of similar 
visual amplitude are 0.3 mag or less.
Furthermore, the location of the pulsation sequences is taken from the LMC
logP-$K$-diagram (Ita et al. 2004), but the actual location will depend on the stellar mass and metallicity.

\section{Conclusions}
From Fourier analysis we could attribute a significant period to each star of our randomly selected sample of irregular variables of type Lb. This period does not completely reproduce the
observed light change, but describes the typical timescale of the variations very well.
There are additional irregularities on top of the light change.

Various parameters derived from Fourier analysis in order to describe the characteristics of the
light change indicate that the irregular variables form a smooth extension of the semiregular
variables towards shorter periods and lower temperature. 
Taking a photometric time series of 1000 to 2000 days length (which is much more than what was typically used to determine the variability class - see Lebzelter et al. 1995) 
there seems to be no difference in the significance of the detected periods between stars classified as irregular or semiregular, respectively, if we take into account differences due to effective temperature
and typical period length. 
We cannot exclude that much longer time series would reveal such a difference. 

It has to be further noted that the same visual light amplitudes are reached in both
cases. Taking into account that the comparison of other stellar parameters like near 
infrared colours or mass loss also show no significant difference between these two variable classes
we have to doubt that {\it Lb} is indeed a distinct and physically meaningful class of variables. 
The Lb stars seem to form the extension of the SRVs towards shorter periods and likely
higher temperatures. This typically corresponds
to smaller light amplitudes. Taking into account the irregularities found on top of the underlying light
change of both period classes it seems well possible that short time series and observational
errors lead to a nondetection of the period in these objects.
Therefore we suggest to combine both groups of variables. A possible term for such a combined
group could be {\it small amplitude red variables (SARVs)} as suggested by Percy et al. (1996)
or Soszynski et al. (2004). Such a 'simplification' of the classification scheme also agrees
with a recent discussion in Eyer \& Mowlavi (2008). 

The reason for the irregularity in the light change found in both variable classes could not
be identified yet. Buchler et al. (2004) find that a low dimensional chaotic pulsation dynamics
arising from the nonlinear interaction of two resonant modes can explain best the origin of
the observed irregularities. Such studies, however, require much longer photometric time series
of these stars as were available for this paper. 

\begin{acknowledgements}
The work of TL was funded by the Austrian Science Fund FWF project P20046-N16. We wish to
acknowledge the support by various students of the lecture "Late Stages of Stellar Evolution" 
in the winter semester 2007. 
Furthermore we want to thank Michael Gruberbauer for fruitful 
discussions concerning frequency analysis. This project made use of the SIMBAD database 
operated at CDS, Strasbourg, France. Special thanks go to Franz Kerschbaum for several
important comments on the manu\-script. Finally we wish to thank the referee, F. van Leeuwen,
for a number helpful hints for improving our discussion of the results.
\end{acknowledgements}

\end{document}